\documentclass[aps,twocolumn,amsmath,amssymb,prd]{revtex4-2}

\def\al{\alpha}
\def\be{\beta}

\def\de{\delta}
\def\ep{\epsilon}

\def\th{\theta}

\def\ph{\phi}

\def\om{\omega}

\def\De{\Delta}

\def\La{\Lambda}

\def\Om{\Omega}

\def\fr#1#2{{{#1}\over{#2}}}
\def\frac#1#2{{\textstyle{{#1}\over{#2}}}}

\def\Re{\hbox{Re}\,}
\def\Im{\hbox{Im}\,}

\def\lsim{\mathrel{\rlap{\lower4pt\hbox{\hskip1pt$\sim$}}
    \raise1pt\hbox{$<$}}}
\def\gsim{\mathrel{\rlap{\lower4pt\hbox{\hskip1pt$\sim$}}
    \raise1pt\hbox{$>$}}}

\def\etal{{\it et al.}}

\def\vev#1{\langle {#1}\rangle}

\def\sqr#1#2{{\vcenter{\vbox{\hrule height.#2pt
         \hbox{\vrule width.#2pt height#1pt \kern#1pt
         \vrule width.#2pt}
         \hrule height.#2pt}}}}

\newcommand{\beq}{\begin{equation}}
\newcommand{\eeq}{\end{equation}}
\newcommand{\bea}{\begin{eqnarray}}
\newcommand{\eea}{\end{eqnarray}}
\newcommand{\rf}[1]{(\ref{#1})}

\newcommand{\bM}{\begin{pmatrix}}
\newcommand{\eM}{\end{pmatrix}}

\def\nn{\nonumber}

\def\f{w}

\def\mbf#1{\boldsymbol #1}

\def\V{\mathcal V}

\def\T{\mathcal T}

\def\K{\mathcal K}

\def\pvec{\mbf p}

\def\bevec{\mbf\be}

\def\pmag{|\pvec|}

\def\punit{\hat p}

\def\phat{\mbf\punit}

\def\nr{{\rm NR}}

\def\nrtemplate#1#2#3{#1^{\nr#3}_{#2}}

\def\cs133{\rm Cs}

\def\Vnrf#1#2{\nrtemplate{{\V_{#1}}}{#2}{}}
\def\TzBnrf#1#2{\nrtemplate{{\T_{#1}}}{#2}{(0B)}}

\def\ToBnrf#1#2{\nrtemplate{{\T_{#1}}}{#2}{(1B)}}

\def\anrf#1#2{\nrtemplate{{a_{#1}}}{#2}{}}
\def\cnrf#1#2{\nrtemplate{{c_{#1}}}{#2}{}}
\def\gzBnrf#1#2{\nrtemplate{{g_{#1}}}{#2}{(0B)}}
\def\goBnrf#1#2{\nrtemplate{{g_{#1}}}{#2}{(1B)}}

\def\HzBnrf#1#2{\nrtemplate{{H_{#1}}}{#2}{(0B)}}
\def\HoBnrf#1#2{\nrtemplate{{H_{#1}}}{#2}{(1B)}}

\def\sVnrf#1#2{\nrtemplate{{\V_{#1}}}{#2}{,{\rm Sun}}}

\def\sTzBnrf#1#2{\nrtemplate{{\T_{#1}}}{#2}{(0B),{\rm Sun}}}
\def\sToBnrf#1#2{\nrtemplate{{\T_{#1}}}{#2}{(1B),\rm Sun}}

\def\widecheck#1{\hskip#1pt\huge$\check{}$}
\def\bighacek#1#2{\vbox{\ialign{##\crcr\widecheck#2\crcr
  \noalign{\kern-9.5pt\nointerlineskip}
   $\hfil\displaystyle{#1}\hfil$\crcr}}}

\def\chM{\vartheta}

\def\TL{T_L}

\begin{document}

\title{Prospects for testing Lorentz symmetry with highly charged ions}

\author{ Arnaldo J.\ Vargas$^1$}

\affiliation{
$^1$Laboratory of Theoretical Physics, Department of Physics, University of Puerto Rico, R\'io Piedras, Puerto Rico 00936\\
}

\begin{abstract}
We investigate the prospects for testing Lorentz symmetry using highly charged ions. We derive the Lorentz-violating energy shifts relevant to Zeeman and Zeeman-hyperfine transitions within the ground state of H-like ions and to Zeeman and Zeeman-fine transitions within the ground state of B-like ions. The resulting signals include sidereal variations through the first harmonic of the sidereal frequency for H-like ions and through the third harmonic for B-like ions. For certain SME coefficients whose contributions scale with the fourth power of the electron momentum, the large electron momentum in H-like ions can provide sensitivity competitive with the best existing bounds at the precision achievable in H-like-ion spectroscopy. We also find that B-like-ion spectroscopy can provide competitive sensitivity to certain SME coefficients.

\end{abstract}
\maketitle

\section{Introduction}

Lorentz symmetry is foundational to both the Standard Model and General Relativity. Lorentz violation has been proposed as a possible low-energy signature of candidate theories of quantum gravity, including string theory \cite{ksp}. This possibility has motivated a broad experimental program aimed at systematically testing Lorentz symmetry, facilitated by the framework provided by the Standard-Model Extension (SME) \cite{sme}. The SME is an effective field theory that extends the conventional Lagrangian densities of the Standard Model and General Relativity by including operators that violate Lorentz symmetry. Each Lorentz-violating operator in the SME Lagrangian density is accompanied by a coefficient that controls the magnitude of the corresponding Lorentz-violating term, conventionally referred to as a coefficient for Lorentz violation or an SME coefficient.

Experimental searches for signals of Lorentz violation can be interpreted as measurements of, or bounds on, combinations of SME coefficients. Definitive evidence for a nonzero SME coefficient would constitute evidence of Lorentz violation. Conversely, an experiment that detects no signal of Lorentz violation at a given sensitivity imposes constraints on the relevant combinations of coefficients. Existing limits obtained from a wide variety of experiments are collected and regularly updated in the {\it Data Tables for Lorentz and CPT Violation} \cite{tables}.

Expressing the results of different tests of Lorentz symmetry in terms of SME coefficients provides a direct means of comparing experiments performed using otherwise very different physical systems. It also makes it possible to identify which experiments provide the strongest sensitivity to particular coefficients, which tests probe complementary combinations of coefficients, and where different measurements provide overlapping sensitivity. In this way, the SME provides both a theoretical description of possible Lorentz violation and a common framework for organizing and comparing the diverse experimental searches for such effects.

In its early development, the SME was restricted to operators of mass dimension $d\leq 4$, since operators of higher mass dimension lead to a nonrenormalizable theory. However, because the SME is intended to describe the low-energy effective field theory that may arise from an underlying theory of quantum gravity, there is no fundamental reason to impose the restriction $d\leq 4$. This restriction was therefore subsequently lifted. For historical reasons, Lorentz-violating operators with mass dimension $d\leq 4$ are referred to as minimal operators, while those with $d\geq 5$ are called nonminimal operators. The systematic classification and incorporation of nonminimal SME coefficients of arbitrary mass dimension remains an active area of research, with results available for several sectors of the SME \cite{km09, km12, km13, nonmingrav, kl21}.  One motivation for studying Lorentz violation in the nonminimal sector, as discussed in Refs.~\cite{km09, km13}, is that in some scenarios describing the emergence of the low-energy effective field theory from an underlying high-energy theory, contributions from nonminimal SME operators to observables are expected to be suppressed at low energies, whereas the corresponding minimal SME contributions need not be similarly suppressed. In scenarios in which such a natural suppression of the nonminimal contributions occurs, and assuming that the possible minimal contributions are absent, this suppression could help explain why evidence for Lorentz violation has remained elusive at currently accessible energy scales. 

Numerous theoretical models for testing Lorentz symmetry through atomic spectroscopy have been developed within the SME framework \cite{that1, that2, that3, that4, that5, that6, that7, that8, that9, that10, that11, that12, that13, that14, gkv14, kv15, kv18, v24, v25, mu25}. These studies have motivated a wide range of experiments searching for Lorentz violation in atomic systems \cite{xema, ph01, hu03, xema2, cs06, al09, Xe09, HeK, NeRb, pe12, ho13, Xe14, pr15, sa19, dr22, oh22, no24, qi25, ba25, na26}. As expected, many of the early models for testing Lorentz symmetry through atomic spectroscopy were restricted to the minimal SME. Refs.~\cite{kv15, kv18} provide comprehensive treatments of how the phenomenology of Lorentz-symmetry tests using atomic spectroscopy changes when nonminimal operators are included. In particular, two features become more prominent in the nonminimal SME. First, some nonminimal contributions exhibit a stronger dependence on the momentum of the fermion within the atom than contributions arising from the minimal SME. Second, the dependence of the contributing SME coefficients on the angular-momentum quantum numbers of the states involved becomes more pronounced in the nonminimal case. Consequently, the atomic spectroscopy experiments that are optimal for constraining minimal SME coefficients are not necessarily those best suited for constraining coefficients in the nonminimal SME.

For instance, within the minimal SME, improving upon the constraints on proton-sector coefficients obtained from the hydrogen-maser experiment \cite{ph01} through deuterium spectroscopy would require comparable or better sensitivity to the Lorentz-violating frequency shift. This is not the case for nonminimal terms \cite{kv15}. The larger momentum of the proton inside the deuteron enhances the sensitivity of deuterium relative to hydrogen to nonminimal proton-sector coefficients \cite{v24}. This enhancement allowed bounds on some nonminimal SME coefficients to be improved by up to 14 orders of magnitude \cite{na26}, even though the deuterium experiment was approximately four orders of magnitude less sensitive to the Lorentz-violating frequency shift and to the leading minimal SME coefficients than the hydrogen-maser experiment.

A similar enhancement associated with a larger momentum has been discussed as an advantage of muonic atoms over muonium, because the momentum of the muon is larger in the former \cite{gkv14}. The physical origin of the enhancement differs between these two examples. For the proton in deuterium, it arises from the proton's motion within the nucleus \cite{kv15, v24}, whereas for the muon in a muonic atom, it arises primarily from the larger reduced mass relative to that of muonium \cite{gkv14}. Neither mechanism is directly available for electrons in ordinary atoms. The present work explores whether a corresponding enhancement in the momentum of valence electrons can instead be obtained using highly charged ions (HCIs), in which these electrons experience a stronger effective Coulomb field. 

A further motivation for this work is the increasing interest in HCIs as systems for precision spectroscopy \cite{ko18}. Their strong electromagnetic fields and large relativistic effects make them particularly suitable for tests of bound-state QED. In addition, several HCIs possess properties that make them promising candidates for optical clocks and other precision tests of fundamental physics. Advances in the trapping, cooling, and spectroscopic interrogation of HCIs have also substantially improved experimental control over these systems and opened new possibilities for high-precision spectroscopy. Continued improvements in HCI spectroscopy, combined with the enhancement arising from the larger momentum of their valence electrons, may therefore make HCIs competitive systems for constraining nonminimal SME coefficients.

In this work, we consider only Zeeman and hyperfine-Zeeman transitions within the ground state of H-like ions. We concentrate on H-like ions because, for a given net ionic charge, their open-shell electron is expected to have the largest momentum. The other class of systems considered consists of Zeeman-fine transitions within the ground-state configuration of B-like ions. The main motivation for considering these systems is that states with total electronic angular momentum $J=3/2$ can receive contributions from SME coefficients that cannot contribute to states with $J=1/2$ \cite{kv15, kv18}. Consequently, the ground-state fine-structure splitting of B-like ions is sensitive to SME coefficients that cannot be studied through transitions within the ground state of H-like ions.

This paper is organized into six sections. Following the Introduction, Sec.~\ref{sec1} presents the general form of the Lorentz-violating energy shift in the laboratory frame. Section~\ref{sec2} describes the transformation, at zeroth order in the boost, of the effective SME coefficients considered in this work from the Sun-centered frame to a laboratory frame on the Earth's surface. The prospects for testing Lorentz symmetry using Zeeman-hyperfine transitions within the ground state of H-like HCIs are discussed in Sec.~\ref{sec3}, while Sec.~\ref{sec4} examines the corresponding prospects using Zeeman-fine transitions within the ground-state configuration of B-like HCIs. The final section summarizes our results and discusses the outlook for future work. Natural units with $\hbar=c=1$ are used throughout.

\section{Lorentz-violating energy shift in the laboratory frame}
\label{sec1}

In this section, we describe the general form of the Lorentz-violating energy shifts used to obtain the corresponding frequency shifts for the Zeeman-fine and Zeeman-hyperfine transitions considered in this work. We consider H-like and B-like ions in the weak-magnetic-field limit, assuming that the total atomic angular momentum quantum number $F$ remains a good quantum number in the presence of an external magnetic field. We use the model presented in \cite{kv18}, which is based on the perturbation derived in \cite{km13}. The general expressions for the Lorentz-violating energy shifts are given in Sec.~II of \cite{kv18}. Applying these results to H-like ions in the weak-magnetic-field limit, we find that the energy shift arising from Lorentz-violating electron operators takes the form
\bea
\de \ep &=& \sum_{kj}\vev{\pmag^k}
\Big(
\Vnrf{e}{kj0}\,\Xi^{(0E)}_j \nn \\
&&+\TzBnrf{e}{kj0}\,\Xi^{(0B)}_j
+\ToBnrf{e}{kj0}\,\Xi^{(1B)}_j
\Big),
\label{AJe}
\eea
where the index $k$ takes the values $0$, $2$, and $4$. The index $j$ takes the values $1$, $3$, and $5$ for the $\T$-type coefficients and $0$, $2$, and $4$ for the $\V$-type coefficients. The expectation values $\vev{\pmag^k}$ correspond to powers of the magnitude of the electron momentum $\pvec$. The coefficients $\Xi$ depend on the angular quantum numbers of the atomic state and can be deduced from the results presented in \cite{kv18}.

For example, using the notation introduced in \cite{kv18}, one may write
\beq
\vev{\pmag^k}\,\Xi^{(0B)}_j=
\vev{F m_F j 0|F m_F}\,
C^{e}_{j}\,
{\La_e}^{(0B)}_{kj}
\label{Xi}
\eeq
with analogous expressions holding for the other $\Xi$ coefficients. Here, $\vev{F m_F j 0|F m_F}$ denotes a Clebsch-Gordan coefficient, $F$ is the total atomic angular momentum, and the expressions for calculating $C^{e}_{j}$ and ${\La_e}^{(0B)}_{kj}$ are given in Eqs. (11) and (18) of \cite{kv18}, respectively.

The coefficients
\bea
\Vnrf{e}{kjm}&=&\cnrf{e}{kjm}-\anrf{e}{kjm},\nn \\
\TzBnrf{e}{kjm}&=&\gzBnrf{e}{kjm}-\HzBnrf{e}{kjm},\nn \\
\ToBnrf{e}{kjm}&=&\goBnrf{e}{kjm}-\HoBnrf{e}{kjm}.
\label{cptnr}
\eea
are the nonrelativistic (NR) coefficients \cite{km13}. The $c$-type and $H$-type coefficients are associated with CPT-even operators, whereas the $a$-type and $g$-type coefficients are associated with CPT-odd operators. In experiments involving ordinary matter, as opposed to antimatter, these coefficients appear in the combinations shown in \rf{cptnr}, which define the $\V$-type and $\T$-type coefficients \cite{kv15, kv18}.

The properties of the NR coefficients are discussed in detail in \cite{km13}. Here, we review only those essential to the present work. The index $k$ denotes the power of the momentum appearing in the Lorentz-violating perturbation. This is evident from \rf{AJe}, in which each coefficient with index $k$ is accompanied by the expectation value $\vev{\pmag^k}$. It also follows from \rf{AJe} that an NR coefficient with index $k$ has mass dimension $1-k$, since the coefficients $\Xi$ are dimensionless.

The NR coefficients are linear combinations of the standard SME coefficients appearing in the SME Lagrangian density \cite{km13}. These combinations receive contributions from coefficients associated with Lorentz-violating operators of arbitrary mass dimension \cite{km13}. For instance, coefficients with $j=3$ can receive contributions from Lorentz-violating operators with mass dimensions $d>5$. This feature arises from expressing the Lorentz-violating perturbation as a power series in $\pmag$ by expanding the free-particle energy in powers of $\pmag/m$, where $m$ denotes the electron mass in the present case. Although no truncation of this expansion is imposed in constructing the general NR perturbation, the resulting expression is referred to as nonrelativistic because the expansion is valid only in the regime $\pmag/m<1$. As a result, SME coefficients associated with operators of different mass dimensions enter through specific linear combinations, with the differences in their mass dimensions compensated by appropriate powers of the electron mass. For convenience, these combinations are defined as the NR coefficients; see Eqs. (111) and (112) of \cite{km13}.

Although the general nonrelativistic perturbation formally contains NR coefficients with arbitrary values of $k$, the analysis leading to \rf{AJe} retains only terms with $k\leq4$ \cite{kv15, kv18}. This truncation includes the complete set of dominant contributions from standard SME operators with mass dimensions up to $d=8$ \cite{kv15}. Operators with $d>8$ can still contribute to the retained NR coefficients, but these contributions are indistinguishable from those of lower-dimensional operators entering the same combinations. However, terms with higher values of $k$ that are required to capture the complete set of dominant effects from operators with $d>8$ are omitted. Consequently, \rf{AJe} is complete at leading order only through mass dimension $d=8$.

The perturbation used to obtain the Lorentz-violating energy shift \rf{AJe} is a single-particle perturbation that describes how Lorentz violation affects the propagation of each fermion within the atom \cite{kv15}. As discussed in \cite{kv18}, each fermion contributes to the Lorentz-violating atomic energy shift through a term of the form given in \rf{AJe}. In this work, however, we consider only the electron contribution and neglect those from the nucleons. Accordingly, the NR coefficients in \rf{AJe} are associated with Lorentz-violating electron operators, as indicated by the subscript $e$. This is also why the mass mentioned in the discussion of the relation between the NR coefficients and the usual SME coefficients is identified with the electron mass, as the present analysis is restricted to the electron sector.

There is a compelling reason to focus on the electron-sector contribution. From \rf{AJe}, we see that the sensitivity to an NR coefficient scales approximately as the ratio between the experimental sensitivity to the Lorentz-violating energy shift and the corresponding momentum expectation value $\vev{\pmag^k}$. As discussed in greater detail in the following sections, although HCI experiments may achieve poorer absolute sensitivity to Lorentz-violating energy shifts than their neutral-atom counterparts, the larger electron momentum in HCIs can compensate for this difference for coefficients with $k=4$. As we show in this work, this enhancement can make HCI experiments more sensitive to these NR coefficients than experiments involving neutral atoms. The momentum-induced enhancement would be even greater for coefficients with $k>4$. Nevertheless, to remain consistent with previous work, we retain contributions only through $k=4$. There is no fundamental obstacle to extending the analysis to include coefficients with $k=6$, thereby obtaining a framework that captures the complete set of leading-order contributions from Lorentz-violating operators with mass dimensions up to $d=10$.

The situation is different for nucleons. Their contribution would be described by an expression similar to \rf{AJe}, with the momentum interpreted as the total momentum of the nucleon in the rest frame of the ion. This momentum includes contributions from both the motion of the nucleus and the nucleon's motion within the nucleus. The internal momenta of nucleons within the nucleus are already large, an effect that has been used, for example, to constrain nucleon NR coefficients through deuterium spectroscopy \cite{v24, na26}. Although the nucleus of the HCI has a considerably larger momentum than in the corresponding neutral system, its contribution to the total momentum of the nucleon is negligible, since the momentum of the nucleon within the nucleus is several orders of magnitude larger. Therefore, there is no significant enhancement of the nucleon momentum that could compensate for the less favorable experimental sensitivity of HCIs to Lorentz-violating energy shifts compared with their neutral counterparts. Consequently, HCIs are not competitive for placing bounds on nucleon coefficients.

We also consider B-like ions. In principle, an energy shift of the form given in \rf{AJe} should be included for each electron in the ion. However, for the transitions considered in this work, the Lorentz-violating shift is dominated by the contribution from the single valence electron, whose energy shift can be described using \rf{AJe}. This follows because the total contribution from electrons in a closed shell vanishes for anisotropic coefficients, namely, coefficients with index $j\ne 0$, leaving only possible contributions from the coefficients with $j=0$, known as isotropic coefficients \cite{kv18}. The contribution of the isotropic coefficients to a transition frequency depends on the differences in the relevant momentum expectation values between the two states. For transitions within the ground-state configuration of a B-like ion, these differences are expected to be small relative to the expectation values themselves. We therefore neglect all isotropic contributions, including those from the core electrons, and retain only the anisotropic contribution from the valence electron.

\section{Laboratory frame to Sun-centered frame transformation of the NR coefficients}
\label{sec2}

The SME coefficients can be interpreted as the components of Lorentz-violating background tensor fields that couple to the usual field operators. As tensor components, they are frame-dependent and transform under changes of reference frame, also known as observer transformations \cite{sme, tables}. While it is natural to express these coefficients as Cartesian components of Lorentz tensors in the SME Lagrangian, in many applications, particularly at the level of the Hamiltonian, it is convenient to express the Lorentz-violating operators as spherical tensors to facilitate the implementation of observer rotations. This procedure can be understood as a spherical decomposition, and the resulting effective coefficients are referred to as spherical coefficients \cite{km13}.

The NR coefficients are spherical coefficients, and the indices $j$ and $m$ specify their transformation properties under observer rotations. A detailed discussion of the rotational properties of the NR coefficients, and more generally of spherical coefficients, can be found in \cite{km13}. The NR coefficients transform as the duals of the spherical harmonics $Y_{jm}(\phat)$ \cite{km13}. The meanings of the indices $j$ and $m$ can therefore be inferred from those of the corresponding indices of the spherical harmonics. A spherical harmonic $Y_{jm}$ transforms as the $m$th component of a spherical tensor of rank $j$. The same interpretation applies to the Lorentz-violating operator associated with the corresponding NR coefficient: the index $j$ specifies its spherical rank, whereas $m$ specifies its spherical component.

For example, the Lorentz-violating operator associated with $\Vnrf{e}{k00}$ transforms as a scalar, the operator associated with $\Vnrf{e}{k1m}$ transforms as a vector, and the operator associated with $\Vnrf{e}{k2m}$ transforms as a rank-2 spherical tensor. Under observer rotations, the coefficients transform as the duals of the corresponding Lorentz-violating operators, ensuring that the perturbation remains invariant. As with spherical harmonics, rotations of the coefficients are implemented using Wigner-$D$ matrices of the corresponding rank $j$ \cite{km13}. The spherical decomposition of the perturbation employs the same conventional spherical basis as the spherical harmonics \cite{km13}, and the resulting NR coefficients are generally complex. However, because the perturbation is Hermitian, coefficients with opposite values of $m$ satisfy the conjugation relation
\beq
({\K_e}_{kjm}^{\rm NR})^*=(-1)^m{\K_e}_{kj(-m)}^{\rm NR}
\label{phR},
\eeq
where ${\K_e}_{kjm}^{\rm NR}$ denotes an arbitrary NR coefficient. This relation implies that a coefficient with $m=-|m|$ is not independent of the corresponding coefficient with $m=|m|$. In this work, we express the observables in terms of the real and imaginary parts of the coefficients with $m>0$, thereby also accounting for all contributions from coefficients with $m<0$. For instance,
\bea
\Re\left[{\K_e}_{kjm}^\nr\right]&=&\dfrac{1}{2}\left({\K_e}_{kjm}^\nr+(-1)^m {\K_e}_{kj(-m)}^\nr\right), \nn\\
\Im\left[{\K_e}_{kjm}^\nr\right]&=&-\dfrac{i}{2}\left({\K_e}_{kjm}^\nr-(-1)^m {\K_e}_{kj(-m)}^\nr\right),
\eea
where $\Re[z]$ and $\Im[z]$ denote the real and imaginary part of $z$, respectively. The coefficients with $m=0$ are real, as follows from \rf{phR}.

Because the SME coefficients are frame-dependent, it is important to report all constraints on SME coefficients in the same reference frame to allow meaningful comparisons of bounds on Lorentz violation across experiments, with the Sun-centered celestial-equatorial frame being the standard frame adopted in the literature for this purpose \cite{tables}. In this frame, the Cartesian coordinates are denoted $(T, X, Y, Z)$, with the origin of the time coordinate $T$ defined at the vernal equinox of the year 2000 and the origin of the spatial coordinates located at the position of the Sun. The $X$ axis points from the location of the Earth at $T = 0$ toward the Sun, and the $Z$ axis is aligned with the Earth's rotation axis.

The usual procedure is to obtain an expression for the Lorentz-violating observable, in this case the Lorentz-violating frequency shift, in the laboratory frame. This is followed by a transformation in which the laboratory-frame SME coefficients appearing in the Lorentz-violating observable are expressed in terms of the Sun-centered-frame SME coefficients. An analysis of the resulting Lorentz-violating signals then yields constraints on the Sun-centered-frame SME coefficients, which constitute the main results of the Lorentz-symmetry test.

The relation between the laboratory-frame and Sun-centered-frame coefficients is obtained by performing an observer Lorentz transformation. The observer Lorentz transformation 
\beq
\La^\mu_{\ \nu}({\mbf\th},\bevec)=
\mathcal{R}^\mu_{\ \al}({\mbf \th})\mathcal{B}^\al_{\ \nu}(\bevec).
\label{lortr}
\eeq
from the Sun-centered frame to the laboratory frame can be obtained by performing a boost $ \mathcal{B}^\mu_{\ \nu}(\bevec)$ with boost parameter $\bevec$, followed by a rotation $ \mathcal{R}^\mu_{\ \nu}({\mbf \th})$ with rotation parameter $\mbf \th$. For a laboratory on the surface of the Earth, $\bevec$ is the velocity of the laboratory relative to the Sun-centered frame which has a small magnitude,  $\be\simeq10^{-4}$, in natural units. This allows to expand the boost transformation $ \mathcal{B}^\mu_{\ \nu}(\bevec)$ as a truncated power series of $\be$ to a desired order. Applying this approximation to linear-boost order in \rf{lortr}, we obtain 
\beq
\La^{0}_{T}=1,
\hskip 8pt
\La^{0}_{J}=-\bevec^J,
\hskip 8pt
\La^{j}_{T}=-\mathcal{R}^{j}_{J}\bevec^{J},
\hskip 8pt
\La^{j}_{J}=\mathcal{R}^{j}_{J},
\label{LTlinear}
\eeq
where lowercase and uppercase indices represent spatial Cartesian coordinates in the laboratory frame and in the Sun-centered frame, respectively. 

Let us concentrate on the transformation at zeroth-boost order. From \rf{LTlinear}, we can deduce that the transformation corresponds to a pure rotation. In this case, the properties of the spherical coefficients become particularly useful because their transformation under rotations is relatively simple. The energy shift \rf{AJe} assumes a laboratory frame with its quantization axis aligned with the external magnetic field, and only laboratory-frame coefficients with index $m=0$ contribute to this shift. Therefore, we need only the expressions for the laboratory-frame $m=0$ coefficients in terms of the Sun-centered-frame coefficients.

We assume that the orientation of the magnetic field is fixed relative to the Earth's surface, so the field rotates with the Earth in the Sun-centered frame. In principle, specifying the complete transformation between the two frames requires defining the other two axes of the laboratory frame. However, this is unnecessary for the laboratory-frame coefficients with $m=0$, which are invariant under rotations about the quantization axis. The relation between a generic laboratory-frame NR coefficient ${\K_e}_{kj0}^{\nr}$ and the corresponding Sun-centered-frame coefficients ${\K_e}_{kjm}^{\rm NR, Sun}$, distinguished by the superscript ${\rm Sun}$, is given by
\beq
{\K_e}_{kj0}^{\rm NR} =
\sum_{m} e^{i m\om_\oplus \TL}
d^{j}_{0m}(-\chM)
{\K_e}^{\rm NR,Sun}_{kjm}.
\label{ltos}
\eeq
Here, $\om_\oplus\simeq 2\pi/(23{\rm ~h} ~56{\rm ~min})$ is the sidereal angular frequency, $\chM$ is the angle between the applied magnetic field and the Earth's rotation axis, and $\TL$ is the local sidereal time, which differs from the Sun-centered-frame time $T$ by an amount that depends on the orientation of the magnetic field. The quantities $d^{j}_{mm'}$ denote the elements of the small Wigner matrices, as given in Eq. (136) of Ref. \cite{km09}.

The angle $\chM$ can be expressed in terms of the local orientation of the magnetic field relative to the laboratory frame and the experiment colatitude $\chi$ as 
\beq
\cos{\chM}= \cos{\th_l} \cos{\chi}+\sin{\th_l} \sin{\chi} \sin{\ph_l},
\label{chM}
\eeq
Here, $\ph_l$ represents the local cardinal direction of the magnetic field, measured counterclockwise from local East. For instance, the local East and local North directions correspond to $\phi_l=0$ and $\ph_l=\pi/2$, respectively. The angle $\th_l$ specifies the orientation of the magnetic field relative to the local vertical direction, where $\th_l=0$ indicates a magnetic field pointing toward the zenith and $\th_l=\pi/2$ indicates a horizontal magnetic field. The local sidereal time $\TL$ is defined with an offset relative to $T$ to absorb a possible phase in the argument of the exponential in \rf{ltos}. Specifically, $T_L=0$ is defined as the moment when the magnetic field is perpendicular to $\hat{Y}$ in the Sun-centered frame and has a nonnegative $X$ component.

The transformation \rf{ltos} shows that the laboratory-frame coefficients are generally time dependent. This is consistent with the noninertial nature of the rotating laboratory frame as the SME coefficients assumed to be constant in an inertial reference frame need not remain constant in a noninertial frame. For the spectroscopy experiments considered here, this variation is sufficiently slow compared with the internal atomic dynamics to justify the adiabatic approximation. The coefficients can also be treated as approximately constant during individual measurements, provided that the measurement duration is short compared with the sidereal period. Nevertheless, their sidereal variation induces corresponding variations in the energy shift \rf{AJe} and the Lorentz-violating frequency shift, which can be study by comparing a sequence of measurements taken at different sidereal times.

All the Lorentz-violating frequency shifts considered in the following sections receive contributions only from the laboratory-frame $\T$-type coefficients with $j=1$ and $j=3$, as well as from the $\V$-type coefficients with $j=2$. The relations between these coefficients and the corresponding Sun-centered-frame coefficients are given by
\begin{widetext}
\bea
\TzBnrf{\f}{k10}&=&\cos{\chM}\,\sTzBnrf{\f}{k10}-\sqrt{2} \sin{\chM}\left(\Re{\left[\sTzBnrf{\f}{k11}\right]}\cos{(\om_\oplus \TL)}-\Im{\left[\sTzBnrf{\f}{k11}\right]}\sin{(\om_\oplus \TL)}\right) \label{sun1}
\eea
\bea
\Vnrf{\f}{k20} &=& \dfrac{1}{4}(1+3\cos{(2\chM)})\sVnrf{\f}{k20}  -\sqrt{\dfrac{3}{2}} \sin{(2 \chM)}\left(\Re{\left[\sVnrf{\f}{k11}\right]}\cos{(\om_\oplus \TL)}-\Im{\left[\sVnrf{\f}{k11}\right]}\sin{(\om_\oplus \TL)}\right)\nn\\
&&+\sqrt{\dfrac{3}{2}} \sin^2{\chM}\left(\Re{\left[\sVnrf{\f}{k22}\right]}\cos{(2\om_\oplus \TL)}-\Im{\left[\sVnrf{\f}{k22}\right]}\sin{(2\om_\oplus \TL)}\right)\label{sun2}
\eea
\bea
\TzBnrf{\f}{k30} &=&  \dfrac{1}{8}(3 \cos{\chM}+5  \cos{(3 \chM)}) \sTzBnrf{\f}{k30}\nn\\
&&-\dfrac{\sqrt{3}}{8} (\sin{\chM}+5 \sin{(3 \chM)} )\left( \Re{\left[\sTzBnrf{\f}{k31}\right]}\cos{(\om_\oplus \TL)}-\Im{\left[\sTzBnrf{\f}{k31}\right]}\sin{(\om_\oplus \TL)}\right)\nn \\
&&+\sqrt{\dfrac{15}{2}} \cos{\chM}\,\sin^2{\chM}\left( \Re{\left[\sTzBnrf{\f}{k32}\right]}\cos{(2 \om_\oplus \TL)} -\Im{\left[\sTzBnrf{\f}{k32}\right]}\sin{(2 \om_\oplus \TL)}\right)\nn \\
&&-\dfrac{\sqrt{5}}{2} \sin^3{\chM}\left( \Re{\left[\sTzBnrf{\f}{k33}\right]}\cos{(3 \om_\oplus \TL)}-\Im{\left[\sTzBnrf{\f}{k33}\right]}\sin{(3 \om_\oplus \TL)}\right)\label{sun3}
\eea
\end{widetext}
with expressions for the transformation of the $\T$-type coefficients with the superscript $1B$ analogous to those for the coefficients with the superscript $0B$. Here, $\Re{[x]}$ is the real part of $x$ and  $\Im{[x]}$ the imaginary part. 

In this work, we do not present explicit expressions for the transformation of the NR coefficients at linear-boost order. A description of how to express the laboratory-frame NR coefficients in terms of SME Sun-centered-frame coefficients up to linear-boost order is provided in \cite{kv18}. Explicit applications of this procedure to the $\T$-type coefficients with $j=1$ can be found in \cite{kv15, kv18}, and corresponding results for the $\V$-type coefficients with $j=2$ are presented in \cite{v24}. No analogous explicit calculation exists for the $\T$-type coefficients with $j=3$. Nevertheless, the general features of the transformation discussed in \cite{kv18}, together with the available explicit examples, allow us to identify the types of Lorentz-violating signals that would arise at linear-boost order in the experimental scenarios considered in the following sections.

\section{Zeeman-Hyperfine and Zeeman transitions in H-like HCI}
\label{sec3}

In this section, we consider the Lorentz-violating frequency shifts of Zeeman-hyperfine transitions within the ground state of H-like ions. All the ions considered have a single electron with orbital angular momentum $L=0$ and total electronic angular momentum $J=1/2$. We focus on the weak-magnetic-field limit, in which the Zeeman shift is assumed to be considerably smaller than the hyperfine splitting, as is typically the case in HCI experiments. For nuclei with nuclear spin $I=0$, we work in the limit in which the magnetic-field-induced splitting is considerably smaller than the fine-structure splitting.

Using \rf{AJe}, we obtain that the Zeeman-hyperfine energy levels within the ground state have the form
\bea
\de \ep&=&2\dfrac{I-F}{2I+1 } \dfrac{m_F}{ \sqrt{3\pi}} (\TzBnrf{\rm e}{010}+2\ToBnrf{\rm e}{010})\nn \\
&&+2\dfrac{I-F}{2I+1 } \dfrac{m_F}{ \sqrt{3\pi}}(\TzBnrf{\rm e}{210}+2\ToBnrf{\rm e}{210})\vev{\pmag^2}\nn\\
&&+2\dfrac{I-F}{2I+1 } \dfrac{m_F}{ \sqrt{3\pi}}(\TzBnrf{\rm e}{410}+2\ToBnrf{\rm e}{410})\vev{\pmag^4},\quad
\label{HFde}
\eea
where $\vec{F}=\vec{I}+\vec{J}$ is the total angular momentum of the ion, and $m_F$ denotes the quantum number associated with its component along the quantization axis. For $I=0$, the formula remains valid under the identifications $F=J$ and $m_F=m_J$. The $\V$-type NR coefficients with $j=0$, known as the isotropic coefficients, also contribute to the Lorentz-violating energy shift. However, their contributions depend only on the principal quantum number and are independent of the angular momentum quantum numbers \cite{kv15}. Consequently, they shift all Zeeman-hyperfine energy levels within the ground state by the same amount and therefore do not contribute to transition frequencies within that manifold. For this reason, we do not include them in \rf{HFde}. They can, however, contribute to gross-structure transitions in which the principal quantum number changes \cite{kv15}.

\renewcommand\arraystretch{1.65}
\begin{table}
\caption{ Expectation values of $\pmag^k$, in units of GeV$^k$, for the ions listed. For $k=4$, the results labeled Dirac were calculated using the DIRAC program, whereas those labeled NR were obtained using the nonrelativistic wave function. For $k=2$, the values were calculated directly from the relativistic wave functions and coincide with those obtained using the DIRAC program.}
\setlength{\tabcolsep}{5pt}
\begin{tabular}{cccc}
\hline
\hline
Ion  & $\vev{\pmag^2}$      &    Dirac $\vev{\pmag^4}$  & NR $\vev{\pmag^4}$ \\ 
&(GeV$^2$) & (GeV$^4$) & (GeV$^4$) \\ \hline
$ {\rm Si}^{13+} $ &  $3\times10^{-9}$ &  $4\times10^{-16}$  & $4\times10^{-17}$ \\
$ {\rm Sn}^{49+} $ &     $5\times10^{-8}$ &  $3\times10^{-13}$ & $6\times10^{-15}$ \\
$ {\rm Cs}^{54+} $ &  $6\times10^{-8}$ &  $6\times10^{-13}$ & $9\times10^{-15}$  \\
$ {\rm Ho}^{66+} $ &  $1\times10^{-7}$ &  $2\times10^{-12}$ & $2\times10^{-14}$  \\
$ {\rm Re}^{74+} $ & $2\times10^{-7}$ &  $5\times10^{-12}$ & $3\times10^{-14}$\\ 
${\rm Tl}^{80+} $ &  $2\times10^{-7}$ &  $1\times10^{-11}$ & $4\times10^{-14}$\\ 
$ {\rm Pb}^{81+} $  &  $2\times10^{-7}$ &  $1\times10^{-11}$ & $4\times10^{-14}$\\ 
${\rm Bi}^{82+} $  &  $2\times10^{-7}$ &  $1\times10^{-11}$ & $5\times10^{-14}$  \\ 
${\rm U}^{91+} $ &   $4\times10^{-7}$ &  $3\times10^{-11}$ & $7\times10^{-14}$  \\\hline
\end{tabular}
\label{table1}
\end{table}

The expectation value $\vev{\pmag^2}$ can be obtained using the H-like solutions of the Dirac Hamiltonian \cite{bd64}. The results for the ions considered are presented in Table \ref{table1} in units of GeV$^2$. By contrast, the expectation value $\vev{\pmag^4}$ calculated using the H-like Dirac Hamiltonian for an $L=0$ state diverges when the nucleus is modeled as a point particle. This divergence can be avoided by modeling the nucleus using a finite charge distribution. For this purpose, we performed Dirac-Hartree-Fock calculations using DIRAC \cite{dirac} with uncontracted Dyall basis sets \cite{dyall} and the standard Gaussian nuclear charge-distribution model implemented in DIRAC \cite{dirac, vd97}. The resulting values of $\vev{\pmag^4}$ are presented in the second-to-last column of Table \ref{table1} in units of GeV$^4$. In the relativistic case, the region in which the electron wave function overlaps the nucleus makes a substantial contribution to $\vev{\pmag^4}$, resulting in a strong dependence on the nuclear model. For this reason, these results are reported to one significant figure. For consistency, the results obtained using the other methods are reported to the same precision. The last column contains the values of $\vev{\pmag^4}$ obtained using the nonrelativistic H-like wave function. Because these values imply weaker sensitivity to the SME coefficients, we use them as a conservative worst-case scenario when estimating the sensitivity of the different HCI experiments to the NR coefficients with $k=4$.

As anticipated in the Introduction and Sec.~\ref{sec1}, the expected advantage of testing Lorentz symmetry using Zeeman-hyperfine transitions within the ground state of H-like HCIs, compared with hydrogen, lies in their enhanced sensitivity to coefficients with $k=4$ due to the higher magnitude of the momentum of the electron. It is therefore worth discussing this point in some detail. We can deduce from \rf{HFde} that the sensitivity to these coefficients is proportional to the ratio of the sensitivity to the Lorentz-violating frequency shift $\de \nu$ to the expectation value $\vev{\pmag^4}$. The best constraints \cite{kv15, tables} on the $\T$-type NR coefficients with $k=4$ were obtained from the results of a hydrogen-maser experiment \cite{ph01} with a sensitivity of approximately 0.4 mHz to a Lorentz-violating frequency shift. The hydrogen-maser experiment studied Zeeman transitions within the ground-state $F=1$ hyperfine manifold. However, because many of the HCI spectroscopy experiments considered in this section aim to measure the ground-state hyperfine transition and would use measurements of this transition to constrain Lorentz violation, it is convenient to express their sensitivity to a Lorentz-violating frequency shift relative to the ground-state hyperfine-transition frequency. Nevertheless, as mentioned previously, the resulting constraints ultimately depend on the absolute sensitivity $\de \nu$ to the Lorentz-violating frequency shift achieved using any Zeeman-hyperfine transition within the ground state, including Zeeman transitions within a single hyperfine level. Apart from angular factors, the attainable sensitivity to the coefficients with $k=4$ is proportional to
\beq
\dfrac{\de \nu}{\vev{\pmag^4}}=\dfrac{\nu}{\vev{\pmag^4}}\times\dfrac{\de \nu}{\nu},
\eeq
where $\nu$ is the ground-state hyperfine-transition frequency. Therefore, for two systems that achieve the same relative frequency uncertainty $\de \nu/\nu$, the difference in their sensitivities to the coefficients is determined by the corresponding values of $\nu/\vev{\pmag^4}$. We can now examine how this ratio varies with the nuclear charge of H-like ions.

In the nonrelativistic limit and within the point-particle nuclear model, both the expectation value $\vev{\pmag^4}$ and the hyperfine energy scale as $Z^4$ \cite{eides}. However, the hyperfine energy is also inversely proportional to the nuclear mass, which generally increases with $Z$, making its effective dependence closer to $Z^3$ than to $Z^4$. Consequently, the ratio $\nu/\vev{\pmag^4}$ scales approximately as $Z^{-1}$. At a fixed relative frequency uncertainty $\de \nu/\nu$, increasing the nuclear charge of an H-like ion should therefore enhance the sensitivity to these coefficients approximately in proportion to $Z$. Thus, a hyperfine-transition experiment using an HCI could achieve sensitivity to the $k=4$ coefficients comparable to that of a corresponding hydrogen experiment with a lower relative precision, or equivalently, a larger relative frequency uncertainty. In the relativistic case, the comparison is further complicated by the divergence of $\vev{\pmag^4}$ for states with $L=0$ when the nucleus is modeled as a point particle.

Before exploring what can be said about the relativistic case, however, it is worth noting that, for the $k=2$ coefficients, $\vev{\pmag^2}$ scales as $Z^2$ in the nonrelativistic limit, whereas the hyperfine energy scales effectively as $Z^3$. Therefore, at a fixed relative frequency uncertainty, the sensitivity to the $k=2$ coefficients decreases with increasing $Z$, and HCIs may consequently not be good candidates for constraining them. For the coefficients with $k=0$, the situation is considerably less favorable than for those with $k=2$, because the sensitivity decreases approximately as $Z^3$ at the same relative frequency precision.

The expectation values $\vev{\pmag^4}$ presented in Table \ref{table1} are enhanced in the relativistic calculations relative to the nonrelativistic results. However, the magnitude of this enhancement depends on the nuclear model used. As mentioned above, we employ the standard Gaussian nuclear charge-distribution model implemented in DIRAC \cite{dirac, vd97}. In the relativistic case, the region in which the $L=0$ electron wave function overlaps the nucleus makes a substantial contribution to $\vev{\pmag^4}$, resulting in a strong dependence on the nuclear model. For this reason, we retain the nonrelativistic result as a conservative lower estimate of $\vev{\pmag^4}$.

\renewcommand\arraystretch{1.65}
\begin{table}
\caption{Sensitivity to the Lorentz-violating frequency shift required for Zeeman-hyperfine transitions in the listed atoms and ions to achieve the same sensitivity to the laboratory-frame coefficients $\TzBnrf{\rm e}{410}$ and $\ToBnrf{\rm e}{410}$ as the hydrogen-maser experiment. }
\setlength{\tabcolsep}{5pt}
\begin{tabular}{ccccc}
\hline
\hline
Ion  & Dirac  & NR   \\ \hline 
$ {\rm H} $                       &  $4\times 10^{-4}\,{\rm Hz}$ & $4\times 10^{-4}\,{\rm Hz}$\\
$ ^{28}{\rm Si}^{13+} $ &  $3\times 10^{2}\,{\rm Hz}$ & $3\times 10^{1}\,{\rm Hz}$  \\
$ ^{118}{\rm Sn}^{49+} $ &  $2\times 10^{5}\,{\rm Hz}$ & $5\times 10^{3}\,{\rm Hz}$\\
$ ^{133}{\rm Cs}^{54+} $ &  $6\times 10^{4}\,{\rm Hz}$ & $9\times 10^{2}\,{\rm Hz}$  \\
$ ^{165}{\rm Ho}^{66+} $ &  $2\times 10^{5}\,{\rm Hz}$ & $2\times 10^{3}\,{\rm Hz}$ \\
$ ^{185}{\rm Re}^{74+} $ &  $7\times 10^{5}\,{\rm Hz}$ & $4\times 10^{3}\,{\rm Hz}$  \\
$ ^{187}{\rm Re}^{74+} $ &  $7\times 10^{5}\,{\rm Hz}$ & $4\times 10^{3}\,{\rm Hz}$  \\
$ ^{203}{\rm Tl}^{80+} $ &  $4\times 10^{6}\,{\rm Hz}$ & $2\times 10^{4}\,{\rm Hz}$   \\
$ ^{205}{\rm Tl}^{80+} $ &  $4\times 10^{6}\,{\rm Hz}$ & $2\times 10^{4}\,{\rm Hz}$   \\
$ ^{207}{\rm Pb}^{81+} $ &  $4\times 10^{6}\,{\rm Hz}$ & $2\times 10^{4}\,{\rm Hz}$  \\
$ ^{208}{\rm Pb}^{81+} $ &  $8\times 10^{6}\,{\rm Hz}$ & $3\times 10^{4}\,{\rm Hz}$  \\
$ ^{207}{\rm Bi}^{82+} $ &  $8\times 10^{5}\,{\rm Hz}$ & $4\times 10^{3}\,{\rm Hz}$  \\
$ ^{208}{\rm Bi}^{82+} $ &  $8\times 10^{5}\,{\rm Hz}$ & $4\times 10^{3}\,{\rm Hz}$  \\
$ ^{209}{\rm Bi}^{82+} $ &  $8\times 10^{5}\,{\rm Hz}$ & $4\times 10^{3}\,{\rm Hz}$   \\
$ ^{238}{\rm U}^{91+} $ &  $2\times 10^{7}\,{\rm Hz}$ & $6\times 10^{4}\,{\rm Hz}$  \\ \hline
\end{tabular}
\label{table2}
\end{table}

Table \ref{table2} explores whether relativistic corrections enhance the sensitivity of HCI experiments to the $\T$-type coefficients with $k=4$. For ions with nuclear spin $I\ne0$, the table considers transitions of the form $(F,m_F)$ to $(F+1,-m_F-1)$, where $F=I-1/2$ is the atomic angular momentum of the lower ground-state hyperfine level. These transitions are chosen because, as can be deduced from \rf{HFde}, their Lorentz-violating frequency shifts are independent of $m_F$. For ions with nuclear spin $I=0$, we instead consider the Zeeman transition with $J=1/2$ and $\De m_J=1$. The first column lists representative examples of H-like ions. These are the same ions considered in Table \ref{table1}, with the addition of neutral hydrogen as the first entry to provide a reference case. For hydrogen, we assume a sensitivity of $0.4$ mHz to the Lorentz-violating frequency shift. The hydrogen-maser experiment studied Zeeman transitions within the $F=1$ hyperfine level of the hydrogen ground state \cite{ph01}, rather than the transition considered in the table. However, the Lorentz-violating frequency shift has the same form for both types of transitions \cite{kv15}.

The second column of Table \ref{table2} presents the sensitivity to a Lorentz-violating frequency shift that spectroscopy experiments involving the corresponding ions would need to achieve to attain the same sensitivity to the laboratory-frame $\T$-type coefficients as the hydrogen-maser experiment. This column was calculated using the values of $\vev{\pmag^4}$ obtained with DIRAC and presented in the second-to-last column of Table \ref{table1}. The third column presents the corresponding results obtained using the nonrelativistic values of $\vev{\pmag^4}$ presented in the last column of Table \ref{table1}. To illustrate the meaning of these results, in the optimistic scenario based on the DIRAC calculations, achieving a sensitivity of 200 kHz to a Lorentz-violating frequency shift in a Zeeman transition within the ground state of $^{118}$Sn$^{49+}$ would yield constraints on the $k=4$ coefficients comparable to those obtained from the hydrogen-maser experiment, which achieved a sensitivity of $0.4$ mHz. In the conservative scenario based on the nonrelativistic calculation, a sensitivity of $5$ kHz would be required. These results provide useful benchmarks for the sensitivity that H-like HCI experiments must achieve to improve the current bounds on the $k=4$ coefficients.

Ground-state hyperfine transitions of H-like HCIs have been measured by emission spectroscopy in electron-beam ion traps \cite{ebit} and by laser spectroscopy of ion beams stored in storage rings \cite{Bi82,Pb81}, whereas Larmor frequencies have been measured in single-ion Penning-trap experiments \cite{si13, sn49}.

The sensitivity required to improve the bounds on the $k=4$ coefficients using storage-ring spectroscopy is beyond that achieved even in the most successful experiments of this type, such as the ${\rm Bi}^{82+}$ experiments. Based on the precision of the hyperfine-transition measurement, we estimate a sensitivity to the Lorentz-violating frequency shift of the order of $10^{10}$ Hz, which is at least five orders of magnitude worse than that required under the most favorable scenario, see Table \ref{table2}. The situation is similar for the emission-spectroscopy experiments. However, the precision of measurements of ground-state hyperfine transitions and Zeeman transitions within the ground-state hyperfine manifold of HCIs is expected to improve significantly in the coming years through the ARTEMIS and SPECTRAP experiments at GSI and the ALPHATRAP experiment at MPIK. For example, based on the precision achieved in the Penning-trap measurements of ${\rm Si}^{13+}$ \cite{si13} and ${\rm Sn}^{49+}$ \cite{sn49}, we estimate that a sensitivity to the Lorentz-violating frequency shift of the order of 10 Hz could be achievable. This sensitivity would make an experiment using ${\rm Sn}^{49+}$ more sensitive to $\T$-type NR coefficients with $k=4$ than the hydrogen-maser experiment by approximately two orders of magnitude, even under the most conservative assumptions in Table \ref{table2}. An experiment using ${\rm Si}^{13+}$ would be competitive in the more conservative scenario and would improve the bounds by one order of magnitude in the most favorable scenario, assuming a sensitivity of 10 Hz to Lorentz-violating frequency shifts. This comparison clearly highlights the potential of HCI spectroscopy experiments to improve the constraints on the NR coefficients with $k=4$.

Before moving forward with the discussion, it is important to note that not all Zeeman-hyperfine transitions within the ground state are sensitive to Lorentz violation through the energy shift \rf{HFde}. For instance, for integer $F$, the transition frequency between states with $m_F=0$ is unaffected by Lorentz violation within the model presented in this work, as can be deduced from \rf{HFde}. Another issue to consider is that the model assumes that the Zeeman levels can be resolved. If the Zeeman levels cannot be resolved, the model must be modified significantly, as discussed in \cite{gkv14, kv15, mu25}.

We have so far compared hydrogen with H-like HCIs. However, it is also useful to compare HCIs with heavy neutral atoms. Core electrons in heavy atoms can possess large momenta, raising the possibility of exploiting this feature to enhance sensitivity to Lorentz violation. Unfortunately, the Lorentz-violating coefficients appearing in the energy shift \rf{HFde} are associated with anisotropic operators, and the total contribution of these operators to the energy shift associated with  closed shell electrons vanishes \cite{kv18}. Only electrons in open shells can produce energy shifts involving the $\T$-type coefficients \cite{kv18}. Because these electrons experience substantial screening of the nuclear charge, their characteristic momenta are generally smaller than that of the electron in a H-like HCI with a comparable nuclear charge.

Another possibility is to consider Li-like HCI experiments \cite{lilike}. In these systems, the general expression for the Lorentz-violating energy shift of the valence electron, in the ground state, would have the form given in \rf{HFde}. For the same isotope and the same experimental sensitivity to the Lorentz-violating energy shift, the H-like ion would generally have an advantage because its electron has a larger momentum than the valence electron in the corresponding Li-like ion. This is why we concentrate on H-like HCIs in this section.

However, a Li-like HCI with a sufficiently larger nuclear charge may be more sensitive to Lorentz-violating effects than an H-like HCI with a considerably smaller nuclear charge. The advantage of the H-like system is most evident when the two charge states of the same isotope are compared. Therefore, in practical situations in which spectroscopy of the H-like ion is not feasible or is subject to significant experimental limitations, whereas spectroscopy of the corresponding Li-like ion is feasible, the Li-like system may represent the better option. The analysis would be similar to that presented in this section, with the principal difference being the calculation of the expectation values of the powers of the electron momentum.

Thus far, the discussion has been limited to the relative sensitivity to the laboratory-frame NR coefficients. However, the coefficients to be constrained are the Sun-centered-frame coefficients, which are related to the laboratory-frame coefficients through \rf{sun1} at zeroth-boost order. For any transition within the ground state, we can deduce from \rf{HFde} and \rf{sun1} that the general form of the frequency shift $\de \nu$ is
\beq
{2\pi}\de \nu = A_{0}+ A_{c}\cos{(\om_\oplus \TL)}+A_{s}\sin{(\om_\oplus \TL)}
\label{sunshift1}
\eeq
The expressions for the amplitudes $A_\xi$ depend on the particular transition under consideration. For instance, if we concentrate on Zeeman transitions for isotopes with $I=0$, as in the cases of $^{118}{\rm Sn}^{49+}$ and $^{28}{\rm Si}^{13+}$, the amplitudes are given by
\bea
A_{0}&=& -\cos{\chM}\dfrac{\De m_J}{ \sqrt{3\pi}} \nn \\
&&\quad\times\sum_{k}\vev{\pmag^k}(\sTzBnrf{\rm e}{k10}+2\sToBnrf{\rm e}{k10}),\nn \\
A_{c}&=&\sin{\chM} \De m_J\sqrt{\dfrac{2}{3\pi}} \nn \\
&&\quad\times\sum_{k}\vev{\pmag^k}\Re[\sTzBnrf{\rm e}{k11}+2\sToBnrf{\rm e}{k11}],\nn \\
A_{s}&=&-\sin{\chM}\De m_J\sqrt{\dfrac{2}{3\pi}} \nn \\
&&\quad\times \sum_{k}\vev{\pmag^k}\Im[\sTzBnrf{\rm e}{k11}+2\sToBnrf{\rm e}{k11}], \nn \\
\label{AHF}
\eea
where $\De m_J=\pm 1$.  The most obvious signal of Lorentz violation arising from \rf{sunshift1} is a sidereal variation of the transition frequency. The predicted variation occurs at the first harmonic of the sidereal frequency. A search for such variations can be used to determine or constrain the amplitudes $A_c$ and $A_s$ and, consequently, the real and imaginary parts of the Sun-centered-frame $\V$-type coefficients with $m=1$. As shown in \rf{AHF}, the sensitivity to these coefficients depends on the relative orientation of the magnetic field and the Earth's rotation axis through the factor $\sin{\chM}$ appearing in $A_c$ and $A_s$. The sensitivity is maximized when the magnetic field is perpendicular to the Earth's rotation axis.

The sidereal-variation signal described by \rf{sunshift1} has the same form as that investigated in the hydrogen-maser experiment \cite{ph01}, which found no evidence of sidereal variations at the level of $0.4$ mHz. This result, the one obtained with the hydrogen maser, was subsequently used to constrain the electron-sector Sun-centered-frame $\V$-type coefficients with $m=1$ \cite{kv15}, and these constraints remain the best available bounds on these coefficients \cite{tables}. Searches for sidereal variations are, however, insensitive to coefficients with $m=0$, such as those contributing to the constant term $A_0$ in \rf{sunshift1}. Consequently, neither the sidereal-variation results from the hydrogen-maser experiment nor analogous studies using HCIs can constrain these coefficients.

As shown in \rf{AHF}, the constant term depends on the relative orientation of the magnetic field and the Earth's rotation axis through the factor $\cos{\chM}$. Measurements performed with different magnetic-field orientations can therefore be used to constrain the $m=0$ coefficients. Such a study has been conducted using hyperfine transitions within the ground state of hydrogen \cite{no24}. The resulting constraint of around 50 Hz on the orientation dependence of the transition frequency yielded the best available bounds on the electron-sector Sun-centered-frame $\V$-type coefficients with $m=0$ \cite{tables}.

The expression for the Lorentz-violating frequency shift in terms of Sun-centered-frame coefficients, \rf{sunshift1}, is valid only at zeroth-boost order. The form of the correction at linear-boost order can be inferred from existing calculations, such as those performed for ground-state Zeeman-hyperfine transitions in a hydrogen maser \cite{kv15} and for a Xe-He comagnetometer \cite{kv18}. The resulting signal would have the form
\bea
2\pi \de \nu
&=&
A_0^{(1)}+A_c^{(1)} \cos\om_\oplus T_L+A_s^{(1)} \sin\om_\oplus T_L
\nn\\
&&
+A_{c2}^{(1)} \cos 2\om_\oplus T_L+A_{s2}^{(1)} \sin 2\om_\oplus T_L\nn \\
&&
+A_C^{(1)} \cos\Om_\oplus T+A_S^{(1)} \sin\Om_\oplus T
\nn\\
&&
+\cos \om_\oplus T_L \left(A_{cC}^{(1)} \cos\Om_\oplus T+A_{cS}^{(1)} \sin\Om_\oplus T\right)
\nn\\
&&
+\sin \om_\oplus T_L \left(A_{sC}^{(1)} \cos\Om_\oplus T+A_{sS}^{(1)} \sin\Om_\oplus T\right),
\nn\\ 
\nn\\
\label{sunshift12}
\eea
where $\Om_\oplus\simeq 2\pi/(365.26 \text{ d})$ is the Earth's orbital angular frequency. Although obtaining explicit expressions for the corresponding amplitudes is beyond the scope of this work, we can highlight the main differences in the variation of the transition frequency arising at linear-boost order. These include annual variations, sidereal variations containing harmonics up to the second harmonic of the sidereal frequency, and mixed terms involving both annual and sidereal variations.

\section{Zeeman-fine transition in B-like highly Charged ions}
\label{sec4}

In this section, we consider the Lorentz-violating frequency shift associated with the ground-state splitting of B-like HCIs\cite{blike, sn45, ar13+}. The motivation for studying these systems is twofold. First, the most precise spectroscopic measurement involving HCIs was performed using Ar$^{13+}$ \cite{ar13+}. Second, the ground-state splitting of B-like ions is sensitive to NR coefficients that cannot be probed in experiments involving ground-state transitions in H-like ions.

For convenience, we concentrate on isotopes with $I=0$. The ground-state configuration of a B-like ion can be described as a single valence electron occupying an open shell with orbital angular momentum $L=1$. This configuration is split into two fine-structure levels with $J=1/2$ and $J=3/2$. Because we consider isotopes with $I=0$, the ground-state configuration exhibits no hyperfine splitting. We assume the presence of a weak magnetic field, as in the preceding sections. The Lorentz-violating energy shift for the Zeeman-fine states belonging to the level with $J=1/2$ is given by
\bea
\de \ep_{1/2}&=&-\dfrac{m_J}{ \sqrt{3\pi}}(\TzBnrf{\rm e}{010}-2\ToBnrf{\rm e}{010})\nn \\
&&-\dfrac{m_J}{ \sqrt{3\pi}}(\TzBnrf{\rm e}{210}-2\ToBnrf{\rm e}{210})\vev{\pmag^2}\nn\\
&&-\dfrac{m_J}{ \sqrt{3\pi}}(\TzBnrf{\rm e}{410}-2\ToBnrf{\rm e}{410})\vev{\pmag^4}.\nn\\
\label{Fde1/2}
\eea
This energy shift is sensitive to the same NR coefficients as the energy shift of the Zeeman-hyperfine levels within the ground state of an H-like ion, given in $\rf{HFde}$, although the coefficients enter through different linear combinations. As in the case of transitions within the ground state of an H-like ion, we neglect contributions from the isotropic NR coefficients with $j=0$, because they cancel in any transition within the ground-state configuration of a B-like ion for the same reasons discussed in connection with \rf{HFde}. For the Zeeman states within the $J=3/2$ manifold, the Lorentz-violating energy shift is given by
\bea
\de \ep_{3/2}&=&-\dfrac{m_J}{ 5\sqrt{3\pi}} (\TzBnrf{\rm e}{010}+4\ToBnrf{\rm e}{010})\nn \\
&&-\dfrac{m_J}{ 5\sqrt{3\pi}}(\TzBnrf{\rm e}{210}+4\ToBnrf{\rm e}{210})\vev{\pmag^2}\nn\\
&&-\dfrac{m_J}{ 5\sqrt{3\pi}}(\TzBnrf{\rm e}{410}+4\ToBnrf{\rm e}{410})\vev{\pmag^4},\nn\\
&&-\dfrac{5-4m_J^2}{ 8\sqrt{5\pi}}(\Vnrf{\rm e}{220}\vev{\pmag^2}+\Vnrf{\rm e}{420}\vev{\pmag^4}),\nn\\
&&+\dfrac{m_J^3-\fr{41}{20} m_J}{\sqrt{7\pi}}(\TzBnrf{\rm e}{230}+\fr{4}{\sqrt{6}}\ToBnrf{\rm e}{230})\vev{\pmag^2}\nn\\
&&+\dfrac{ m_J^3-\fr{41}{20} m_J}{\sqrt{7\pi}}(\TzBnrf{\rm e}{430}+\fr{4}{\sqrt{6}}\ToBnrf{\rm e}{430})\vev{\pmag^4}.\nn\\
\label{Fde3/2}
\eea
The Lorentz-violating energy shifts of the $J=3/2$ states are sensitive to $\T$-type coefficients with $j=3$ and $V$-type coefficients with $j=2$, which do not contribute to the energy shifts associated with the ground-state hyperfine structure of hydrogen. The appearance of these coefficients results from the larger electronic angular momentum. As discussed in detail in \cite{kv15, kv18}, an electron Lorentz-violating operator associated with a spherical coefficient with index $j$ can contribute to the energy shift only for states satisfying $J\geq j/2$. Consequently, coefficients with $j=1$ can contribute to states with $J\geq1/2$, whereas, for coefficients with $j=2$ and $j=3$, the smallest allowed half-integer value is $J=3/2$. For this reason, studies of Lorentz violation using transitions within the ground-state configuration of B-like ions are sensitive to a broader set of coefficients than analogous studies using H-like ions.

We used DIRAC \cite{dirac} to calculate $\vev{\pmag^k}$, where $\pmag$ denotes the magnitude of the momentum of the valence electron. The results are presented in Table \ref{table3} for some ions of interest listed in the first column. The values for $k=2$ and $k=4$ are presented in the second and third columns, respectively.
\renewcommand\arraystretch{1.65}
\begin{table}
\caption{ Expectation values of $\pmag^k$, in units of GeV$^k$, for the valence electrons of the listed B-like ions, calculated using the DIRAC program.}
\setlength{\tabcolsep}{5pt}
\begin{tabular}{cccc}
\hline
\hline
Ion  & $\vev{\pmag^2}$      &    Dirac $\vev{\pmag^4}$   \\ 
&(GeV$^2$) & (GeV$^4$)  \\ \hline
$ {\rm S}^{11+} $ &  $9\times10^{-10}$ &  $2\times10^{-18}$  \\
$ {\rm Cl}^{12+} $ &     $1\times10^{-9}$ &  $2\times10^{-18}$  \\
$ {\rm Ar}^{13+} $ &  $1\times10^{-9}$ &  $3\times10^{-18}$  \\
$ {\rm K}^{14+} $ &  $1\times10^{-9}$ &  $4\times10^{-18}$  \\
$ {\rm Sn}^{45+} $ &  $9\times10^{-9}$ &  $2\times10^{-16}$  \\
$ {\rm Pb}^{77+} $ &  $3\times10^{-8}$ &  $2\times10^{-15}$  \\ \hline\hline
\end{tabular}
\label{table3}
\end{table}

Comparing the results in Tables \ref{table1} and \ref{table3}, we find that, even for ions with similar ionic charges, the expectation values $\vev{\pmag^k}$ are significantly larger for the electron in an H-like ion than for the valence electron in a B-like ion. The B-like valence electron occupies a higher principal shell and is therefore, on average, farther from the nucleus than the electron in the corresponding H-like ion. In addition, its $p$-type orbital has less overlap with the region near and within the nuclear charge distribution than the $s$-type orbital of the H-like ion. Together, these effects reduce the momentum of the B-like valence electron and, consequently, the expectation values $\vev{\pmag^k}$, with the reduced nuclear overlap being particularly important for $\vev{\pmag^4}$.

Therefore, experimental studies of ground-state splittings in H-like ions generally have greater sensitivity to $\T$-type coefficients with $j=1$ than analogous studies using B-like ions with similar nuclear charges. The advantage of B-like-ion experiments is that the $J=3/2$ manifold provides sensitivity to coefficients with $j>1$ that cannot be accessed through studies of ground-state splittings in H-like ions. However, Zeeman transitions within the $J=1/2$ manifold of a B-like ion cannot generally compete with equivalent transitions in an H-like ion of similar nuclear charge. Both systems probe the same set of coefficients, but the smaller momentum expectation values of the B-like valence electron reduce its sensitivity. For instance, although Zeeman transitions within the $J=1/2$ manifold of the ground-state configuration of the B-like ion $^{118}{\rm Sn}^{45+}$ \cite{sn45} have been measured with an absolute frequency precision similar to that achieved for the H-like ion $^{118}{\rm Sn}^{49+}$ \cite{sn49}, the latter is considerably more sensitive to Lorentz violation.

Before discussing the best current bounds on the NR coefficients with $j>1$, it is useful to express the frequency shift in terms of the Sun-centered-frame coefficients. We concentrate on the $^2P_{1/2} {-} ^2P_{3/2}$ transitions within the ground-state configuration of a B-like ion. Furthermore, we neglect contributions from the $j=1$ coefficients because they can be probed more effectively using H-like ions. Nevertheless, the form of the contribution from the $j=1$ coefficients to the frequency shift can be deduced by inspection of \rf{AHF} and by comparing the energy shifts in the B-like and H-like cases. Using \rf{sun2} and \rf{sun3} together with \rf{Fde3/2}, we find that the Lorentz-violating frequency shift in terms of the Sun-centered-frame coefficients has the form
\bea
{2\pi}\de \nu&=&   A_{0}+A_{c}\cos{(\om_\oplus \TL)}+A_{s}\sin{(\om_\oplus \TL)}\nn\\
&&+ A_{c2}\cos{(2\om_\oplus \TL)}+A_{s2}\sin{(2\om_\oplus \TL)}\nn\\
&&+ A_{c3}\cos{(3\om_\oplus \TL)}+A_{s3}\sin{(3\om_\oplus \TL)},
\label{sunshift2}
\eea
where the amplitudes $A_\xi$ are given in Table \ref{table4}. In this table, the quantum number $m_j$ corresponds to the state with $J=3/2$, and the sum over $k$ runs over the values $k=2$ and $k=4$. The result is independent of the quantum number $m_J$ of the state with $J=1/2$ because we have neglected contributions from the $j=1$ coefficients.  

\renewcommand{\arraystretch}{3}
\begin{table*}
\caption{Expressions for the amplitudes appearing in \rf{sunshift2}. The NR coefficients entering the coefficient combinations are expressed in the Sun-centered frame, and the quantum number $m_j$ labels the Zeeman level of the $P_{3/2}$ state. } \setlength{\tabcolsep}{5pt} \begin{tabular}{cl} \hline
\hline																												
	$A_\xi$     &			Coefficient  combination	 	\\	\hline
$	A_0		          $  &$\displaystyle \sum_k\left(-\dfrac{5-4m_J^2}{ 32\sqrt{5\pi}}(1+3\cos{2\chM})\sVnrf{e}{k20}+\dfrac{m_J^3-\fr{41}{20} m_J}{8 \sqrt{7\pi}}(3 \cos{\chM}+5  \cos{3 \chM}) \big(\TzBnrf{\rm e}{k30}+\fr{4}{\sqrt{6}}\ToBnrf{\rm e}{k30}\big)\right)\vev{\pmag^k}$\nn \\
$	A_c	          $  &$ \displaystyle\sqrt{3}\sum_k\left(\dfrac{5-4m_J^2}{ 8\sqrt{10\pi}}\sin{2\chM}\,\Re{\left[\sVnrf{e}{k20}\right]}-\dfrac{m_J^3-\fr{41}{21} m_J}{8 \sqrt{7\pi}}(\sin{\chM}+5 \sin{3 \chM} ) \Re\left[ \TzBnrf{\rm e}{k31}+\fr{4}{\sqrt{6}}\ToBnrf{\rm e}{k31}\right]\right)\vev{\pmag^k}$\nn \\
$	A_s	          $  &$\displaystyle-\sqrt{3}\sum_k\left(\dfrac{5-4m_J^2}{ 8\sqrt{10\pi}}\sin{2\chM}\,\Im{\left[\sVnrf{e}{k20}\right]}-\dfrac{m_J^3-\fr{41}{21} m_J}{8 \sqrt{7\pi}}(\sin{\chM}+5 \sin{3 \chM} ) \Im\left[ \TzBnrf{\rm e}{k31}+\fr{4}{\sqrt{6}}\ToBnrf{\rm e}{k31}\right]\right)\vev{\pmag^k}$\nn \\
$	A_{c2}	          $  &$ \displaystyle-\sqrt{3}\sum_k\left(\dfrac{5-4m_J^2}{ 8\sqrt{10\pi}}\sin^2{\chM}\,\Re{\left[\sVnrf{e}{k22}\right]}-\dfrac{m_J^3-\fr{41}{20} m_J}{\sqrt{14\pi}}\sqrt{5}\,\cos{\chM}\,\sin^2{\chM} \,\Re\left[ \TzBnrf{\rm e}{k32}+\fr{4}{\sqrt{6}}\ToBnrf{\rm e}{k32}\right]\right)\vev{\pmag^k}$\nn \\
$	A_{s2}	          $  &$ \displaystyle\sqrt{3}\sum_k\left(\dfrac{5-4m_J^2}{ 8\sqrt{10\pi}}\sin^2{\chM}\,\Im{\left[\sVnrf{e}{k22}\right]}-\dfrac{m_J^3-\fr{41}{20} m_J}{\sqrt{14\pi}}\sqrt{5}\,\cos{\chM}\,\sin^2{\chM} \,\Im\left[ \TzBnrf{\rm e}{k32}+\fr{4}{\sqrt{6}}\ToBnrf{\rm e}{k32}\right]\right)\vev{\pmag^k}$\nn \\
$	A_{c3}	          $  & $\displaystyle-\dfrac{m_J^3-\fr{41}{20} m_J}{2\sqrt{7\pi}}\sqrt{5}\,\sin^3{\chM} \sum_k\Re\left[ \TzBnrf{\rm e}{k32}+\fr{4}{\sqrt{6}}\ToBnrf{\rm e}{k33}\right]\vev{\pmag^k}$\nn \\
$	A_{s3}	          $  & $\displaystyle\dfrac{m_J^3-\fr{41}{20} m_J}{2\sqrt{7\pi}}\sqrt{5}\,\sin^3{\chM} \sum_k\Im\left[ \TzBnrf{\rm e}{k32}+\fr{4}{\sqrt{6}}\ToBnrf{\rm e}{k33}\right]\vev{\pmag^k}$\nn \\
\hline
\hline
\end{tabular}
\label{table4}
\end{table*} 

Table \ref{table4} illustrates several general features of the sidereal-variation signal expressed in terms of the Sun-centered-frame NR coefficients at zeroth order in the boost. First, coefficients with a given value of $|m|$ contribute only to the amplitude of the corresponding $|m|$th harmonic of the sidereal frequency in \rf{sunshift2}. This general feature is discussed and explained in \cite{kv15, kv18}. Second, because $j\geq |m|$, a coefficient with index $j$ can contribute only up to the $j$th harmonic of the sidereal variation of the transition frequency. Consequently, the $\V$-type coefficients with $j=2$ can contribute only up to the second harmonic, whereas the $\T$-type coefficients with $j=3$ can contribute up to the third harmonic of the sidereal frequency.

The Sun-centered-frame electron-sector NR coefficients with $j>1$ and $m\ne0$ remain unconstrained. These coefficients generate sidereal variations of the transition frequency, whereas those with $m=0$ contribute only to a constant shift. By contrast, nucleon-sector NR coefficients with $j=2$ and $m\ne0$ have been constrained using the results of sidereal-variation studies \cite{kv18, na26}. The coefficients with $j=3$ and $m\ne0$, however, remain unconstrained in both the electron and nucleon sectors.

The situation differs for the NR coefficients with $m=0$, which contribute to the constant term $A_0$ in \rf{sunshift2}. Bounds on the $V$-type coefficients with $j=2$ and $m=0$ and the $\T$-type coefficients with $j=3$ and $m=0$ have been obtained by comparing the measured $1S-2P$ transition frequency in antihydrogen with the corresponding prediction for hydrogen \cite{1s2p, v25}. The methods used to constrain coefficients with $m=0$ are diverse and can be classified into three broad categories: comparisons between theoretical predictions and experimental measurements \cite{gkv14, kv15, oh22}, searches for variations of the transition frequency with the orientation of the magnetic field relative to the Earth's rotation axis \cite{no24}, and comparisons between matter and antimatter systems \cite{kv18, v25, ba25}. In the context of HCIs, the second approach appears particularly promising because it does not require an absolute theoretical prediction for the transition frequency. By contrast, constraints derived from direct comparisons between theory and experiment will be limited by the largest between the uncertainty of the theretical prediction or experimental result \cite{gkv14, kv15}. 

The current bounds on the NR coefficients with $j>1$ and $m=0$ were derived from measurements of the $1S-2P$ transition in antihydrogen, for which an absolute uncertainty of the order of 100 MHz was achieved \cite{1s2p}. This uncertainty is considerably larger than those achieved for the principal transitions targeted in antihydrogen spectroscopy, namely, the $1S-2S$ transition \cite{ah18} and the ground-state hyperfine splitting \cite{ak26}, whose absolute uncertainties have reached the kilohertz range. Both transitions, however, involve only electronic states with $J=1/2$ and are therefore insensitive to coefficients with $j>1$, as discussed previously. HCI experiments consequently offer favorable prospects for improving the existing bounds, particularly because their sensitivity is enhanced by the larger characteristic momentum of the valence electron relative to that in hydrogen.

HCI experiments could provide the first constraints on NR coefficients with $j>1$ and $m\ne0$. To evaluate their prospects for constraining these coefficients, we consider the $^2P_{1/2}-^2P_{3/2}$ transition in Ar$^{13+}$, for which a fractional systematic uncertainty of the order of $10^{-17}$ has been achieved \cite{ar13+}. To suppress systematic effects, the experiment averaged four transitions
\bea
 &&m_J=1/2\rightarrow m_J=1/2,\nn\\
 &&m_J=1/2\rightarrow m_J=3/2,\nn\\
 &&m_J=-1/2\rightarrow m_J=-1/2\nn\\ 
  &&m_J=-1/2\rightarrow m_J=-3/2.
\eea
Averaging a pair of transitions related by $m_J\rightarrow-m_J$, a Zeeman pair, cancels the linear Zeeman shift. For example, the average of the $m_J=1/2\rightarrow m_J=3/2$ and $m_J=-1/2\rightarrow m_J=-3/2$ transition frequencies is insensitive to the linear Zeeman shift. This averaging also cancels contributions from the $\T$-type coefficients with $j=3$, which are odd functions of $m_J$, while retaining sensitivity to the $\V$-type coefficients with $j=2$, whose contributions are even functions of $m_J$.

Further averaging the two Zeeman-pair frequencies, which is equivalent to averaging all four transition frequencies, cancels both the electric-quadrupole shift and the contribution from the $\V$-type coefficients with $j=2$ appearing in \rf{Fde3/2}. This cancellation occurs because the electric-quadrupole interaction transforms as a spherical tensor of rank $j=2$, and averaging over all four $m_J$ sublevels of the $J=3/2$ manifold cancels the contribution from any spherical tensor of this rank, including the Lorentz-violating operators associated with the $\V$-type $j=2$ coefficients. Consequently, within the model considered here, the four-transition average used as the clock frequency is insensitive to both the $\V$-type coefficients with $j=2$ and the $\T$-type coefficients with $j=3$.

The quadrupole shift of an individual transition, in the Ar$^{13+}$ experiment, was estimated to be approximately 10 mHz, below the statistical uncertainty of approximately 100 mHz \cite{ar13+}. Therefore, an analysis based only on the individual Zeeman-pair averages could retain the high precision of the Ar$^{13+}$ experiment. The advantage of averaging over only one of the Zeeman pairs is that it retains sensitivity to the $\V$-type coefficients with $j=2$, for which our model predicts sidereal variations up to the second harmonic of the sidereal frequency, see Table \ref{table4}. A sidereal-variation search could compare one of the Zeeman-pair averages with the four-transition average, which, within the model considered here, should not contain any Lorentz-violating sidereal variation and could therefore serve as a reference frequency.

Although the charge of Ar$^{13+}$ is not particularly large, the characteristic momentum of its valence electron is substantially enhanced relative to that in hydrogen. However, this momentum may not be significantly larger than those of valence electrons in heavy neutral atoms or heavy ions with lower charge states. For instance, the clock transitions used in the $^{171}$Yb$^+$ and $^{88}$Sr$^+$ optical ion clocks are sensitive to NR coefficients with $j>1$ before the averaging procedures used to cancel the linear Zeeman and electric-quadrupole shifts are applied \cite{kv18}. By contrast, the Sr and Yb optical lattice clocks are sensitive only to NR coefficients with $j=0$ \cite{kv18}. The frequency of the ground-state fine-structure transition in Ar$^{13+}$ \cite{ar13+} is comparable to the clock-transition frequencies of the Sr$^+$ and Yb$^+$ ion clocks \cite{optical}, all of which are of the order of $10^{14}$ Hz.

Among the electronic states involved in these clock transitions, only the valence-electron states $F_{7/2}$ in Yb$^+$ and $D_{5/2}$ in Sr$^+$ have $J>1/2$ and therefore receive Lorentz-violating energy shifts involving NR coefficients with $j>1$. According to our calculations performed using DIRAC, for a valence electron in the state $F_{7/2}$ in Yb$^+$, we obtain 
\bea
\vev{\pmag^2}&\simeq& 7\times 10^{-10}\,{\rm GeV}^2,\nn\\
 \vev{\pmag^4}&\simeq& 1\times 10^{-18}\, {\rm GeV}^4.
 \eea
For a valence electron in the state $D_{5/2}$ in Sr$^+$, we obtain 
\bea
\vev{\pmag^2}&\simeq& 4\times 10^{-11}\,{\rm GeV}^2,\nn\\
 \vev{\pmag^4}&\simeq& 4\times 10^{-20}\, {\rm GeV}^4.
 \eea
Comparing these values with the results in Table \ref{table3}, and noting that the three systems have transition frequencies of the same order, we find that Ar$^{13+}$ and Yb$^+$ should exhibit sensitivities of approximately the same order to the NR coefficients at the same relative precision, up to differences in their angular factors. By comparison, Ar$^{13+}$ benefits from the larger characteristic momentum of its valence electron relative to that in Sr$^+$, resulting in an enhancement in sensitivity of approximately one order of magnitude. Nevertheless, Sr$^+$ and Yb$^+$ currently appear to have an advantage because of their lower fractional systematic uncertainties, which have reached the $10^{-18}$ level for Yb$^+$ and the $10^{-19}$ level for Sr$^+$ \cite{optical}, compared with approximately $10^{-17}$ for Ar$^{13+}$ \cite{ar13+}. However, the precision of measurements of the Ar$^{13+}$ ground-state splitting may improve substantially \cite{ar13+}, potentially making this system competitive with Sr$^+$ and Yb$^+$.

For ions with higher charge states, such as Sn$^{45+}$, achieving sensitivity competitive with that of the Sr$^+$ and Yb$^+$ clocks may nevertheless require a fractional precision comparable to that required for Ar$^{13+}$ for the $k=4$ coefficients. In the nonrelativistic limit, both the fine-structure splitting and $\vev{\pmag^4}$ scale approximately as the fourth power of the effective nuclear charge \cite{eides}. Consequently, the enhancement in $\vev{\pmag^4}$ with increasing ionic charge is accompanied by a comparable increase in the transition frequency, so the relative precision required to attain a given sensitivity to the $k=4$ coefficients does not necessarily improve substantially. In contrast to the hyperfine splitting, the dominant contribution to the fine-structure splitting is not inversely proportional to the nuclear mass. Therefore, the increase in nuclear mass with increasing nuclear charge does not weaken the effective fourth-power scaling of the fine-structure splitting. As in the case of the ground-state hyperfine splitting of H-like ions discussed previously, the situation is less favorable for coefficients with $k=2$. Because $\vev{\pmag^2}$ scales approximately as the square of the effective charge while the fine-structure splitting scales approximately as its fourth power, progressively higher relative precision is required as the ionic charge increases.

\section{Outlook and Summary}
In this work, we explored the prospects for testing Lorentz symmetry using highly charged ions. Section~\ref{sec1} presents the general form of the Lorentz-violating energy shift in the laboratory frame, given in \rf{AJe}. Section~\ref{sec2} discusses how the laboratory-frame coefficients can be expressed in terms of coefficients in the Sun-centered frame at zeroth-boost order, with the explicit transformations relevant to this work given in \rf{sun1}, \rf{sun2}, and \rf{sun3}.

In Sec.~\ref{sec3}, we considered Zeeman and Zeeman-hyperfine transitions within the ground state of H-like ions, whose Lorentz-violating energy shifts are given in \rf{HFde}. We examined the advantages offered by H-like HCIs for constraining the $\T$-type NR coefficients with $k=4$ and showed that the precision already achieved in HCI spectroscopy could improve the existing bounds on these coefficients. In particular, this conclusion follows from the uncertainty achieved in a recent Penning-trap measurement of Sn$^{49+}$ \cite{sn49}. The corresponding frequency shift in terms of the Sun-centered-frame coefficients is given in \rf{sunshift1} and predicts sidereal variations at the first harmonic of the sidereal frequency. We also discussed strategies for constraining the coefficients with $m=0$, which do not contribute to sidereal variations at zeroth-boost order.

In Sec.~\ref{sec4}, we considered Zeeman and Zeeman-fine transitions within the ground-state configuration of B-like ions. As can be inferred by comparing Tables \ref{table1} and \ref{table3}, the enhancement associated with the larger electron momentum is less pronounced for B-like ions than for H-like ions. The Lorentz-violating energy shifts of the $J=1/2$ and $J=3/2$ states are presented in \rf{Fde1/2} and \rf{Fde3/2}, respectively. Despite their smaller momentum enhancement, B-like ions offer the important advantage that the $J=3/2$ states are sensitive to NR coefficients with $j=2$ and $j=3$, which do not contribute to any energy level within the ground state of an H-like ion. Our analysis indicates that the recent high-precision measurement of the ground-state fine-structure splitting in Ar$^{13+}$ \cite{ar13+} is close to being competitive with leading optical clocks \cite{optical} in sensitivity to anisotropic NR coefficients with $k=4$. This conclusion applies to the individual Zeeman-fine transitions rather than to the averaged clock frequency, because the averaging scheme used to cancel the linear Zeeman and electric-quadrupole shifts also cancels the Lorentz-violating contributions considered here. However, an alternative scheme that cancels only the linear Zeeman shift while retaining the electric-quadrupole shift may preserve both high precision and sensitivity to the NR coefficients with $j=2$. Overall, HCIs offer promising and potentially competitive opportunities for testing Lorentz symmetry, particularly through searches involving the $k=4$ coefficients.

The expectation values of powers of the electron momentum used in this work were evaluated by post-processing DHF wave functions obtained with the DIRAC program. The property-analysis tools provided by DIRAC were used to calculate both the total expectation values and the contributions from individual orbitals. For H-like ions, for which electron correlation is absent, the principal sources of uncertainty include the nuclear charge-distribution model and possible basis-set incompleteness, particularly in the description of the short-distance and high-momentum behavior relevant to $\vev{\pmag^4}$. For B-like ions, additional uncertainty arises from correlation effects beyond the DHF approximation. Although these effects are generally suppressed in highly charged few-electron ions, they need not be negligible for expectation values of one-particle operators. The uncertainties are expected to be larger for the singly charged ions used in optical clocks, for which electron-correlation effects are stronger and the electronic structures are more complex. Accordingly, the values presented here should be regarded as order-of-magnitude estimates and refined through more detailed calculations in future work.

The Lorentz-violating perturbation considered in this work does not include corrections to the interactions of electrons with external fields or with one another. These terms are expected to be suppressed relative to the dominant contributions retained here \cite{kv15}. Their inclusion also presents a practical challenge because interaction-dependent terms are considerably more complicated to incorporate.

One possible approach would be to use the Lorentz-violating perturbation derived in \cite{dk16}, which describes a charged spin-$1/2$ fermion interacting with external fields and includes contributions from Lorentz-violating operators of mass dimension up to six. This perturbation was developed in the context of searches for Lorentz and CPT violation using Penning traps. Its direct application to many-electron atoms is not straightforward, however, because it describes interactions with external fields rather than interactions among charged particles. Although the electric field produced by the nucleus could be treated as an external field acting on the electrons, electron-electron interactions would require a more complete treatment.

For H-like ions, which contain only one electron, the perturbation of \cite{dk16} could be applied by treating the nuclear electric field as external. The strong nuclear electric field sampled by the bound electron could enhance sensitivity to SME coefficients that couple the fermion field to the electromagnetic field. Investigating these effects therefore represents a particularly interesting extension of the present analysis.

\section*{Acknowledgments} 
We extend our gratitude to Fabian Hei{\ss}e for suggesting exploring the prospects of highly charged ion for testing Lorentz and CPT symmetry and for his insightful comments and valuable feedback.  This work benefited in part from support provided by the National Science Foundation under Award No.~PHY-2607562.

\end{document}